\documentstyle[prc,aps]{revtex}
\begin{document}
\draft
\title{ Nuclear matter properties and relativistic mean-field theory}
\author{K.C. Chung$^1$, C.S. Wang$^{1,2}$, A.J. Santiago$^1$, and 
J.W. Zhang$^2$}

\address{\it (1) Instituto de F\'\i sica, Universidade do Estado do Rio 
de Janeiro,\\
Rio de Janeiro-RJ 20559-900, Brazil \\
(2) Department of Technical Physics, Peking University,
Beijing 100871, China}
\date{\today}
\maketitle
\begin{abstract}
Nuclear matter properties are calculated in the relativistic mean field 
theory by using a number of different parameter sets. The result shows 
that the volume energy $a_1$ and the symmetry energy $J$ are around the 
acceptable values $16$MeV and $30$MeV respectively; the 
incompressibility $K_0$ is unacceptably high in the linear model, but 
assumes reasonable value if nonlinear terms are included; the density 
symmetry $L$ is around $100MeV$ for most parameter sets, and the 
symmetry incompressibility $K_s$ has positive sign which is opposite to 
expectations based on the nonrelativistic model. In almost all parameter 
sets there exists a critical point $(\rho_c, \delta_c)$, where the 
minimum and the maximum of the equation of state are coincident and the 
incompressibility equals zero, falling into ranges 
$0.014$fm$^{-3}<\rho_c<0.039$fm$^{-3}$ and $0.74<\delta_c\le0.95$; for 
a few parameter sets there is no critical point and the pure neutron 
matter is predicted to be bound. The maximum mass $M_{NS}$ of neutron 
stars is predicted in the range 
$2.45$M$_\odot\leq M_{NS}\leq 3.26$M$_\odot$, the corresponding neutron 
star radius $R_{NS}$ is in the range $12.2$km$\leq R_{NS}\leq 15.1$km.
\end{abstract}


\vskip 0.5cm
\centerline{\bf 1. Introduction }
\vskip 0.5cm

Groundstate nuclear matter properties are specified by  the nuclear 
matter equation of state $e(\rho_N,\delta)$ which is simply the energy 
per nucleon of nuclear matter given as a function of nucleon density 
$\rho_N$ and relative neutron excess $\delta=(\rho_n-\rho_p)/\rho_N$. 
This equation of state is a fundamental quantity in theories of neutron 
stars and supernova explosions, as well as in theories of nucleus-nucleus 
collisions at energies where nuclear compressibility comes into 
play\cite{Li}. The main measured quantities which can provide information 
about equation of state are the binding energies and other data from 
finite nuclei. As the finite nuclei are in states near the nuclear matter 
standard state $(\rho_N=\rho_0,\delta=0)$, which is defined as the 
equilibrium state of symmetric nuclear matter with minimum energy per 
nucleon and called also the normal state, our actual knowledge of 
nuclear matter is mainly about nuclear matter at state close to the 
point $(\rho_0, 0)$. In this case, the equation of state can be written 
approximately as\cite{Myers69}\cite{CWSZ00}
\begin{equation}
\label{EoSa}
 e(\rho_N,\delta)=-a_1+\frac 1{18}\big(K_0+K_s\delta^2\big)
 \Big(\frac{\rho_N-\rho_0}{\rho_0}\Big)^2+\Big[J+\frac L3
 \Big(\frac{\rho_N-\rho_0}{\rho_0}\Big)\Big]\delta^2,
\end{equation}
which is specified by the standard density $\rho_0$, volume energy 
$a_1$, symmetry energy $J$, incompressibility $K_0$, density symmetry 
$L$ and symmetry incompressibility $K_s$. The most interesting quantity 
for supernova explosion calculation is the nuclear incompressibility 
$K_0$ which dictates the balance between gravity and internal pressure 
of the stellar system, while the most interesting quantities for heavy- 
ion collision studies are the nuclear incompressibility $K_0$ and the 
symmetry incompressibility $K_s$ which influence the  side-flow effects 
and the isotopic distributions of the collisions, respectively.

There is no direct experimental measurement on these quantities. They 
can be determined only from data fit based on some specific nuclear 
model. Therefore, our actual knowledge about these quantities is 
essentially model dependent. Nowadays the quantities which are known 
with reasonable precision are $a_1$, $J$ and $K_0$, being the last two 
still under active investigation. One of the most sophisticated data fit 
is given by the nonrelativistic Thomas-Fermi statistical model of nuclei 
with Myers-Swiatecki phenomenological nucleon-nucleon 
interaction\cite{Myers96}. It is a fit to 1654 ground-state masses of 
nuclei with $N, Z\ge 8$, together with a constraint that ensures 
agreement with measured values of the nuclear surface diffuseness, 
giving the root-mean-square mass deviation equal to $0.655MeV$. The data 
fits based on Skyrme nucleon-nucleon interactions give comparable  
results\cite{CWSZ00}, whereas a model independent but approximate data 
fit also gives $a_1$, $K_0$, $J$ and $L$ very close to that obtained by 
the before mentioned data fit\cite{CWS99}\cite{WCS99}.

As the $\sigma$-$\omega$-$\rho$ model of the relativistic mean-field 
theory is used widely to investigate various nuclear phenomena with 
success\cite{Serot86}-\cite{Reinhard89}, it is interesting to calculate 
these nuclear matter quantities within this model by using the available 
parameter sets, to compare with those obtained by the nonrelativistic 
model. In addition, as these parameters are determined by nuclear 
ground-state properties, it is also interesting to see what the 
$\sigma$-$\omega$-$\rho$ model can predict for the nuclear system under 
extreme conditions of density and asymmetry. In this case, the most 
interesting quantities are the location $e_m=e(\rho_m,\delta)$ of the 
minimum of the equation of state for given asymmetry $\delta$, and the 
generalized incompressibility $K_m=K(\rho_m,\delta)$ of the nuclear 
matter at this state\cite{Myers98}. Another interesting quantity is the 
maximum mass of neutron stars $M_{NS}$ calculated by the equation of 
state for neutron matter with $\delta=1$. Actually, to predict these 
properties of nuclear matter under extreme conditions is just one of the 
main goals in developing a relativistic mean-field theory\cite{Serot97}.

The purpose of this paper is to make  above mentioned calculation in 
comparing with results obtained by the nonrelativistic model. Section II 
presents the formalism and formulas used in this calculation. Section III 
addresses a numerical analysis on linear $\sigma$-$\omega$-$\rho$ model 
of the relativistic mean-field theory. The standard nuclear matter 
properties calculated from a number of parameter sets are given in 
Section IV, and the prediction for cold nuclear matter under extreme 
conditions is made in Section V. Section VI gives the summary. Appendix 
A displays functions $F_m(x)$ and $f_m(x)$ which are useful in the 
analytical expressions as well as in the numerical calculations. The 
Bjorken-Drell convention for four-vector\cite{Bjorken} and the natural 
units with $\hbar=c=1$ are used.

\vskip 0.5cm
\centerline{\bf 2. Formalism }
\vskip 0.5cm

The $\sigma$-$\omega$-$\rho$ model of the relativistic mean-field theory 
is specified by the following Lagrangian density\cite{Serot97}:
$${\cal L}=\overline\psi[\gamma_\mu(i\partial^\mu
-g_\omega\omega^\mu-g_\rho\mbox{\boldmath $\tau\cdot$}{\bf b}^\mu)
-(M-g_\sigma\phi)]\psi$$
$$+\frac 12(\partial_\mu\phi\partial^\mu\phi
-m_\sigma^2\phi^2)-\frac 13Mb(g_\sigma\phi)^3
-\frac 14c(g_\sigma\phi)^4$$
$$-\frac 14F_{\mu\nu}F^{\mu\nu}+\frac 12m_\omega^2\omega_\mu\omega^\mu
+\frac 14c_3(\omega_\mu\omega^\mu)^2$$
\begin{equation}\label{sorL}
-\frac 14{\bf B}_{\mu\nu}\mbox{\boldmath $\cdot$} 
{\bf B}^{\mu\nu}+\frac 12m_\rho^2{\bf b}_\mu\mbox{\boldmath $\cdot$}
{\bf b}^\mu,
\end{equation}
where $F^{\mu\nu}=\partial^\mu\omega^\nu-\partial^\nu\omega^\mu,$
${\bf B}^{\mu\nu}=\partial^\mu{\bf b}^\nu-\partial^\nu{\bf b}^\mu$, 
$\psi$, $\phi$, $\omega$ and ${\bf b}^\mu$ are the nucleon, $\sigma$, 
$\omega$ and $\rho$ meson fields with masses $M$, $m_\sigma$, $m_\omega$ 
and $m_\rho$, respectively, while $g_\sigma$, $g_\omega$ and $g_\rho$ 
are the respective coupling constants; $b$, $c$ and $c_3$ are the 
nonlinear term coefficients, and {\boldmath $\tau$} are isospin 
matrices. The nuclear matter equation of state derived from this 
Lagrangian density can be expressed in terms of the nuclear energy 
density ${\cal E}$ as $e={\cal E}/\rho_N-M$, and
\begin{equation}\label{Eksor}{\cal E}={\cal E}_k
+{\cal E}_\sigma+{\cal E}_\omega+{\cal E}_\rho,
\end{equation}
\begin{equation}\label{Ek}{\cal E}_k
=\frac{M^4\xi^4}{\pi^2}\sum_{i=p,n}F_1(k_i/\xi M),
\end{equation}
\begin{equation}\label{Esigma}{\cal E}_\sigma
=M^4\Big[\frac 1{2C_\sigma^2}(1-\xi)^2
+\frac 13b(1-\xi)^3+\frac 14c(1-\xi)^4\Big],
\end{equation}
\begin{equation}\label{Eomega}{\cal E}_\omega
=\frac{C_\omega^2\rho_N^2}{2M^2}
\frac 1{(1+c_3\omega_0^2/m_\omega^2)^2}
+c_3\frac 34\frac{C_\omega^4\rho_N^4}{M^4m_\omega^4}
\frac 1{(1+c_3\omega_0^2/m_\omega^2)^4},
\end{equation}
\begin{equation}\label{Erho}{\cal E}_\rho
=\frac{C_\rho^2\rho_N^2}{2M^2}\delta^2,
\end{equation}
where $k_p$ and $k_n$ are the proton and neutron Fermi 
momenta respectively, 
\begin{equation}\label{xi}\xi=\frac{M^*}M
=1-\frac{g_\sigma}M\phi,
\end{equation}
\begin{equation}\label{Ci}C_i=g_i\frac M{m_i},\,\,\,\,
i=\sigma, \omega, \rho, 
\end{equation}
and the function $F_m(x)$ is defined as (see Appendix A for details):
\begin{equation}
F_m(x)=\int_0^xdx\,x^{2m}\sqrt{1+x^2}.
\end{equation}
The reduced effective nucleon mass $\xi$ and thus the field $\phi$ is 
determined by
\begin{equation}\label{Eqxi}(1-\xi)+bC_\sigma^2(1-\xi)^2
+cC_\sigma^2(1-\xi)^3=\frac{C_\sigma^2}{\pi^2}\xi^3\sum_{i=p,n}
f_1(k_i/\xi M),
\end{equation}
and the field $\omega_0$ by
\begin{equation}\label{omega0}
\omega_0=\frac{C_\omega\rho_N}{Mm_\omega}
\frac 1{1+c_3\omega_0^2/m_\omega^2}.
\end{equation}

Knowing the equation of state, the following formulas for 
pressure $p$ and generalized incompressibility $K$\cite{Myers98} can be 
obtained:
\begin{equation}\label{pressure}p=-{\cal E}
+\rho_N\frac{\partial{\cal E}}{\partial\rho_N}
=\frac 13{\cal E}_k-\frac 13M\xi\rho_s-{\cal E}_\sigma
+{\cal E}_\omega-\frac 12c_3\omega_0^4+{\cal E}_\rho,
\end{equation}
$$K\equiv 9\frac{\partial p}{\partial\rho_N}
=\frac 1{\rho_N}\left\{\frac{M^4\xi^4}{\pi^2}
\sum_{i=p,n}\Big(\frac{k_i}{\xi M}\Big)^3f_1'(k_i/\xi M)
+9\frac{C_\rho^2\rho_N^2}{M^2}\delta^2\right.$$
\begin{equation}\label{K}
\left.+9\frac{C_\omega^2\rho_N^2}{M^2}
\frac 1{1+3c_3\omega_0^2/m_\omega^2}
+3\frac{M^4\xi^4}{\pi^2}\sum_{i=p,n}\frac{k_i}{\xi M}
f_1'(k_i/\xi M)
\frac{\rho_N}\xi\frac{\partial\xi}{\partial\rho_N}\right\},
\end{equation}
\begin{equation}\frac{\rho_N}\xi
\frac{\partial\xi}{\partial\rho_N}=\frac 13\frac Q{\xi[1+2bC_\sigma^2(1
-\xi)+3cC_\sigma^2(1-\xi)^2]+Q+3C_\sigma^2\rho_s/M^3},
\end{equation}
\begin{equation}\label{rhos}\rho_s=\frac{M^3\xi^3}{\pi^2}
\sum_{i=p,n}f_1(k_i/\xi M),
\end{equation}
\begin{equation}Q=-\frac{C_\sigma^2}{\pi^2}\xi^3
\sum_{i=p.n}\frac{k_i}{\xi M}f_1'(k_i/\xi M).
\end{equation}
In the previous equation, $f_m'(x)=df_m(x)/dx$, and the function $f_m(x)$ is defined as 
(see Appendix A for details)
\begin{equation}
f_m(x)=\int_0^xdx\frac{x^{2m}}{\sqrt{1+x^2}}.
\end{equation}

At the standard state $(\rho_0, 0)$, the pressure should be 
zero, 
\begin{equation}\label{SC}p\,(\rho_0, 0)=0,
\end{equation}
and 
\begin{equation}\label{K0}K_0=K(\rho_0, 0)=
9\Big(\rho_N^2\frac{\partial^2e}{\partial\rho_N^2}\Big)_0.
\end{equation}
In addition, the following formulas can be derived:
\begin{equation}\label{J}
J\equiv\frac 12\frac{\partial^2e}{\partial\delta^2}\Big|_0
=\frac 16\frac{k_F^2}{\sqrt{k_F^2+M^2\xi_0^2}}
+\frac{C_\rho^2k_F^3}{3\pi^2M^2},
\end{equation}
\begin{equation}\label{L}L\equiv\frac 32\Big(\rho_N
\frac{\partial^3e}{\partial\rho_N\partial\delta^2}\Big)_0
=J+2J_\rho-\left\{3+\frac{M^3}{C_\sigma^2\rho_s}\xi[1
+2bC_\sigma^2(1-\xi)+3cC_\sigma^2(1-\xi)^2]\right\}_0
J_\sigma,
\end{equation}
\begin{equation}\label{Jrho}
J_\rho=\frac{C_\rho^2\rho_0}{2M^2},
\end{equation}
\begin{equation}\label{Jsigma}
J_\sigma=-\frac 32M\Big(\frac{M\xi_0}{k_F}\Big)^3
f_1(k_F/\xi_0 M)\frac{\partial^2\xi}{\partial\delta^2}\Big|_0,
\end{equation}
$$\frac{\partial^2\xi}{\partial\delta^2}\Big|_0=\frac 29
\frac{C_\sigma^2\xi_0^3}{\pi^2}\Big[2\frac k{\xi M}
f_1'(k/\xi M)-\Big(\frac k{\xi M}\Big)^2f_1''(k/\xi M)\Big]_0
$$
\begin{equation}\times\left\{\frac{2C_\sigma^2\xi_0^2}{\pi^2}
\Big[3f_1(k/\xi M)-\frac k{\xi M}f_1'(k/\xi M)\Big]+[1
+2bC_\sigma^2(1-\xi)+3cC_\sigma^2(1-\xi)^2]\right\}_0^{-1},
\end{equation}
\begin{equation}\label{Ks}K_s\equiv\frac 92\Big(\rho_N^2
\frac{\partial^4e}{\partial\rho_N^2\partial\delta^2}\Big)_0
=-6L+\frac 12\frac{\partial^2K}{\partial\delta^2}\Big|_0.
\end{equation}
The subscript $0$ in the above formulas stands for the standard state 
$(\rho_0, 0)$, and $k_F$ is the nucleon Fermi momentum of standard 
nuclear matter which is related to standard density $\rho_0$ and nuclear 
radius constant $r_0$ as
\begin{equation}\label{rho0r0kF}
\rho_0=\frac 1{4\pi r_0^3/3}=\frac{2k_F^3}{3\pi^2}.
\end{equation}

Formula (\ref{J}) is well-known in the literature\cite{Serot86}. It is 
worthwhile to note that, for the linear model with $b=c=c_3=0$, the 
$\rho$-meson in the standard state is nonrelevant to Eqs.(\ref{Eksor}), 
(\ref{Eqxi}) and (\ref{SC}). Thus the parameters $C_\sigma$ and 
$C_\omega$ are the same both for model with or without $\rho$-meson, 
since they are determined by standard density $\rho_0$ and volume energy 
$a_1$. This point will be discussed more specifically in the next 
Section.

\vskip 0.5cm
\centerline{\bf 3. Determination of $C_\sigma$, $C_\omega$ and $C_\rho$ 
in the linear model }
\vskip 0.5cm

For the linear model, $b=c=c_3=0$, the nuclear energy density 
(\ref{Eksor}) in the standard state $(\rho_0,0)$ is simplified as
\begin{equation}\label{Eksorl0}
{\cal E}_0=2\frac{M^4\xi_0^4}{\pi^2}F_1(k_F/\xi_0M)
+\frac{M^4}{2C_\sigma^2}(1-\xi_0)^2
+\frac{C_\omega^2\rho_0^2}{2M^2},
\end{equation}
and Eq.(\ref{Eqxi}) determining the reduced effective nucleon mass $\xi$ 
becomes
\begin{equation}\label{xil0}
1-\xi_0=2\frac{C_\sigma^2\xi_0^3}{\pi^2}f_1(k_F/\xi_0M).
\end{equation}
In addition, the equilibrium condition (\ref{SC}) and the expression of 
incompressibility (\ref{K0}) are reduced, respectively, to
\begin{equation}\label{SCl0}
\frac 23 \frac{M^4\xi_0^4}{\pi^2}f_2(k_F/\xi_0M)
-\frac{M^4}{2C_\sigma^2}(1-\xi_0)^2
+\frac{C_\omega^2\rho_0^2}{2M^2}=0,
\end{equation}
\begin{equation}\label{K0l}
K_0=6\frac{C_\omega^2k_F^3}{\pi^2M^2}
+3M\xi_0f_1'(k_F/\xi_0M)\Big[1+\frac{M^2\xi_0^2}{k_F^2}
\Big(\frac{Q_0}{3-2\xi_0+Q_0}\Big)\Big],
\end{equation}
\begin{equation}\label{Q0}
Q_0=-2\frac{C_\sigma^2\xi_0^3}{\pi^2}\frac{k_F}{\xi_0M}
f_1'(k_F/\xi_0M).
\end{equation}
It can be seen from Eqs.(\ref{Eksorl0})-(\ref{Q0}) that the relevant 
quantities are $k_F$, $C_\sigma^2$, $C_\omega^2$ and $M$. Note that 
$\xi_0$ is determined by Eq.(\ref{xil0}) and $\rho_0$ is related to 
$k_F$ by Eq.(\ref{rho0r0kF}). Therefore, as the measured nucleon mass 
can be taken for $M$, the composite parameters $C_\sigma^2$ and 
$C_\omega^2$ can be determined completely by the value 
$e_0=e(\rho_0,0)$, by using Eqs.(\ref{Eksorl0})-(\ref{SCl0}) together 
with ${\cal E}_0=(e_0+M)\rho_0$. The procedure is as follows.

At the stable equilibrium point $(\rho_0, 0)$, an equation involving 
$e_0$, $k_F$, $\xi_0$ and $C_\sigma$ can be obtained if 
Eqs.(\ref{Eksorl0}) and (\ref{SCl0}) are combined to cancel $C_\omega$. 
On the other hand, $C_\sigma$ can be solved as a function of $k_F$ and 
$\xi_0$ from Eq.(\ref{xil0}). Substituting this function of $C_\sigma$ 
into the above-mentioned equation, the following equation involving 
$e_0$, $k_F$ and $\xi_0$ can be derived:
\begin{equation}\label{Eqxi0}\frac 3{\xi_0}f_1(k_F/\xi_0M)
+2f_2(k_F/\xi_0M)=(e_0+M)\frac{k_F^3}{\xi_0^4M^4}.
\end{equation}
$\xi_0$ can be calculated from this equation, if the location 
$(\rho_0, e_0)$ of stable equilibrium point is chosen as input data. 
Having this $\xi_0$ together with $\rho_0$ and $e_0$, $C_\sigma$ can be 
calculated from Eq.(\ref{xil0}), then $C_\omega$ can be determined from 
Eq.(\ref{Eksorl0}) or (\ref{SCl0}). Finally, the incompressibility $K_0$ 
can be obtained from Eq.(\ref{K0l}). It can be shown easily that 
Eq.(\ref{Eqxi0}) is identical to Eq.(22) of Ref.\cite{Serot97} which is 
originally given in Ref.\cite{Furnstahl}. Numerically, $\rho_0$ and thus 
$k_F$ can be expressed in terms of the nuclear radius constant $r_0$ as 
Eq.(\ref{rho0r0kF}), while $e_0$ can be related to the nuclear volume 
energy coefficient $a_1$ as $e_0=e(\rho_0,0)=-a_1$. The experimentally 
acceptable values are\cite{CWSZ00}
\begin{equation}r_0\approx 1.14{\rm fm},\,\,\,\,
a_1\approx 16{\rm MeV}.
\end{equation}

The numerical calculation shows that, in the ranges 
$1.05$fm$\le r_0\le 1.25$fm and $15.5$MeV$\le a_1\le 16.5$MeV, the 
effective mass $\xi\approx 0.54$ does not depend on the choice of $r_0$ 
and $a_1$ sensitively. The composite parameters $C_\sigma^2$ and 
$C_\omega^2$ are sensitive to the choice of $r_0$ but not of $a_1$. 
Fig.1 shows $C_\sigma^2$ and $C_\omega^2$ as a function of $r_0$ for given 
$a_1=16$MeV. Fig.2 gives the nuclear matter incompressibility $K_0$ 
calculated by Eq.(\ref{K0l}) as a function of $a_1$ for given 
$r_0=1.14$fm. It is not sensitive to the choice of $r_0$. Furthermore, 
Fig.2 shows that $K_0$ is approximately a linear function of $a_1$, in 
agreement with what is obtained in the macroscopic phenomenological 
approach to the nuclear matter\cite{CWSZ1}.

In case of the Walecka model\cite{Walecka74}, $C_\rho=0$, other nuclear 
matter properties $J$, $L$ and $K_s$ can be calculated also from these 
$C_\sigma$ and $C_\omega$ by Eqs.(\ref{J})-(\ref{Ks}). The calculated 
coefficients $J$, $L$ and $K_s$ are almost constant in the range 
$15.5$MeV$\le a_1\le 16.5$MeV for given $r_0=1.14$fm,
\begin{equation}J\approx 20{\rm MeV},\,\,\,\,L\approx 70{\rm MeV},\,\,\,
K_s\approx 88{\rm MeV}.
\end{equation}
On the other hand, these coefficients depend on the choice of 
$r_0$ weakly, for given $a_1$.

In case $\rho$-meson is included also in the model, the composite 
parameter $C_\rho$ can be determined by measured symmetry energy $J$ 
through Eq.(\ref{J}). The inclusion of $\rho$-meson contributes to the 
symmetry energy with an extra term $J_\rho$ (Eq.(\ref{Jrho})) and to the 
density symmetry $L$ with an extra term $3J_\rho$, while keeping the other 
coefficients $a_1$, $K_0$ and $K_s$ unchanged. For symmetry 
incompressibility $K_s$, it can be seen from Eqs.(\ref{Ks}) and 
(\ref{K}) that the $\rho$-meson contributes with a term $-18J_\rho$ to $-6L$ 
and a term $18J_\rho$ to $(1/2)\partial^2K/\partial\delta^2|_0$, and 
these extra terms cancel each other.

\vskip 0.5cm
\centerline{\bf 4. Standard state nuclear matter properties }
\vskip 0.5cm

There are many parameter sets for the $\sigma$-$\omega$-$\rho$ model of the 
relativistic mean field theory in the literature, some of them are 
listed in Table I, where L-W is taken from the pioneering Walecka linear 
$\sigma$-$\omega$ model\cite{Walecka74}, L-HS from the Horowitz-Serot 
linear $\sigma$-$\omega$-$\rho$ model\cite{HS81}, L1, L2 and L3 from 
Lee {\it et al}.\cite{Lee86}, L-Z, NL-Z and NL-VT from 
Rufa {\it et al}.\cite{Rufa88}, NL1 from Reinhard 
{\it et al}.\cite{Reinhard86}, NL2 from Fink {\it et al}.\cite{Fink88}, 
NL3 and NL3-II from Lalazissis {\it et al}.\cite{Lalazissis}, NLB, NLC 
and NLD from Serot\cite{Serot92}, NL-B1 and NL-B2 from 
Boussy {\it et al}.\cite{Boussy84a}\cite{Boussy84b}, NL-RA from 
Rashdan\cite{Rashdan97}, NL-SH from Sharma {\it et al}.\cite{Sharma93}, 
TM1 and TM2 from Sugahara and Toki\cite{Sugahara94}. Most of them are 
collected in Reinhard's review\cite{Reinhard89}. In Table I, $g_2$ and 
$g_3$ are defined, respectively, as
\begin{equation}g_2=Mbg_\sigma^3,\,\,\,\,g_3=cg_\sigma^4.
\end{equation}
It should be noted that some of these parameter sets are given 
originally in values of $C_i$ instead of $g_i$, $i=\sigma$, $\omega$, 
$\rho$. In this case the values of $g_i$ given here are calculated from 
$C_i$, $m_i$ and $M$ by Eq.(\ref{Ci}). 
It should be noted also that our $g_\rho$ is only one half of that 
defined in Ref.\cite{Serot97}.

As mainly nuclear matter properties are concerned in the present 
calculation, the relevant parameters are only $C_\sigma^2$, 
$C_\omega^2$, $C_\rho^2$, while the meson masses $m_\sigma$, $m_\omega$ 
and $m_\rho$ are nonrelevant ones, in case of the linear model. However, 
in nonlinear model the meson mass is able in some case to influence the 
nuclear matter property. For example, the $\omega$ meson mass $m_\omega$ 
appears in Eq.(\ref{omega0}) and thus has effect on nuclear matter 
property in the nonlinear model via the term $(\omega_\mu\omega^\mu)^2$.

The standard nuclear matter properties related to these parameter sets 
are shown in Table II, where all quantities are given in MeV, except 
$\rho_0$ which is in fm$^{-3}$. In the calculation of $a_1$, $K_0$, $J$, 
$L$ and $K_s$, using formulas given in Section II and input parameters 
listed in Table I, Eqs.(\ref{Eqxi}) and (\ref{SC}) should be solved 
simultaneously at first for $\xi_0$ and $k_F$ at the standard point. The 
calculation of $\partial^2K/\partial\delta^2|_0$, in Eq.(\ref{Ks}) of 
$K_s$, is made numerically, as its analytical expression is too 
complicated to be derived. The simple numerical average among the 
nonlinear model sets is given as the set $\langle$NL$\rangle$, and the 
Myers-Swiatecki's result\cite{Myers96} is shown also as the set MS for 
comparison. 

$\rho_0$ and $a_1$ give the location of nuclear matter standard state. 
Most values of $\rho_0$ given in the $\sigma$-$\omega$-$\rho$ model are lower 
than that of Myers-Swiatecki's, the later corresponds to $r_0=1.140$fm 
and agrees with that obtained from elastic electron scattering and 
muonic atom spectroscopy measurements\cite{deVries}\cite{Fricke}.
Most values of $a_1$ given in the $\sigma$-$\omega$-$\rho$ model are in the 
reasonable range around $16$MeV, except those of L1, L2, L3, LZ and NL2 
sets, which seem too large. Since $a_1$ is the leading term in the 
approximate equation of state (\ref{EoSa}), it is the main parameter in 
any data fit to nuclear masses. However, there is a big fluctuation 
around $16$MeV, as can be seen from Table II.

$K_0$ and $J$, the next terms to the leading $a_1$ in the approximate 
equation of state (\ref{EoSa}), are the fine tune in the data fit to 
nuclear masses, as shown in the droplet model of nuclei\cite{Myers77}. 
It can be seen from Table II that $K_0$ given in the 
$\sigma$-$\omega$-$\rho$ model is much larger than that of 
Myers-Swiatecki's, while $J$ is only about $2/3$ of Myers-Swiatecki's, 
for the linear $\sigma$-$\omega$ model; $J$ will be increased if the $\rho$- 
meson is added also to the linear $\sigma$-$\omega$ model, but $K_0$ 
keeps the same value. This is an inherent character of linear 
$\sigma$-$\omega$-$\rho$ model, as has been shown generally in last 
Section. In this respect, the nonlinear terms are needed in order to 
reduce the nuclear incompressibility $K_0$, as supported by the 
calculated results listed in Table II. It is worthwhile to note that, 
even the value of $K_0$ obtained from different nuclear measurements and 
astrophysical observations are spread over a large range from $180$ to 
$800$MeV\cite{Norman}, most expectations based on the nonrelativistic 
model are around $220$MeV\cite{Myers98}.

Being terms of order higher than $K_0$ and $J$ in the approximate  equation of 
state (\ref{EoSa}), $L$ and $K_s$ belong to the superfine tune in the 
data fit to nuclear masses. Even if most values of $L$ given in the 
$\sigma$-$\omega$-$\rho$ model seem to be larger than the acceptable one, 
they are still in the reasonable range around $100MeV$. On the other 
hand, the values of $K_s$ are all positive whose sign is opposite to 
most expectations based on the nonrelativistic model\cite{Li}. 
Experimentally, $K_s$ obtained from the isoscalar giant-monopole 
resonance energy is between $-566\pm 1350$ to 
$34\pm 159$MeV\cite{Shlomo}. 

\vskip 0.5cm
\centerline{\bf 5. Prediction for cold nuclear matter under extreme 
conditions }
\vskip 0.5cm

The stability condition for the state at minimum of equation of state 
for given asymmetry $\delta$ is
\begin{equation}\label{SCdelta}p\,(\rho_m,\delta)=0.
\end{equation}
The solution of this equation for given $\delta$ gives the location of 
the minimum $\rho_m=\rho_m(\delta)$. Knowing this location 
$\rho_m(\delta)$, the minimum $e_m=e(\rho_m,\delta)$ and the generalized 
incompressibility at this minimum $K_m(\delta)=K(\rho_m,\delta)$ can be 
calculated. Furthermore, the critical point of the equation of state 
$(\rho_c,\delta_c)$ can be defined as the point where the maximum and 
the minimum are coincident and thus the curvature of $e(\rho,\delta_c)$ 
{\it versus} $\rho$ equals zero. As the generalized incompressibility 
$K(\rho,\delta)$ is proportional to this curvature, we have at the 
critical point
\begin{equation}K_m(\delta_c)=K(\rho_c,\delta_c)=0.
\end{equation}
This equation together with (\ref{SCdelta}) can be used to obtain the 
critical point $(\rho_c,\delta_c)$.

Table III lists the calculated critical point $(\rho_c,\delta_c)$, the 
corresponding effective nucleon mass $M^*/M$, the energy per nucleon 
$e_m$ as well as the generalized incompressibility $K_m$ at the critical 
point. In case there is no critical point, the corresponding quantities 
at the minimum point of the pure neutron matter equation of state with 
$\delta=1$ are listed. $\rho_c$ is in fm$^{-3}$ units, while $e_m$ and 
$K_m$ are in MeV units. The values given by the Myers-Swiatecki equation of 
state\cite{Myers98} are also listed in the last row for comparison. It 
can be seen that there is no critical point for parameter sets LW, L1, 
L2, L3 and NL-B2. In these cases, there is a minimum for the pure neutron 
matter equation of state and the bound neutron matter is predicted. For 
other parameter sets, the neutron matter is an unbound gas system. The 
predicted critical point $(\rho_c,\delta_c)$ is in the ranges 
$0.014$fm$^{-3}<\rho_c\le0.039$fm$^{-3}$ and $0.74<\delta_c\le0.95$, 
with the corresponding effective nucleon mass in the range 
$0.87\le M^*/M\le0.95$.

In addition, the predicted maximum mass $M_{NS}$ and the corresponding 
radius $R_{NS}$ of neutron stars, calculated by the Oppenheimer-Volkoff 
equation, using the $\sigma$-$\omega$-$\rho$ model equation of state of the 
relativistic mean field theory with the above mentioned parameter sets 
and $\delta=1$, are also shown in Table III. The range of the maximum 
mass is $2.45$M$_\odot\leq M_{NS}\leq 3.26$M$_\odot$, and the range of 
corresponding star radius is $12.2$km$\leq R_{NS}\leq 15.1$km.

Fig.3 gives some examples of $\rho_m(\delta)$, where the solid curve 
from top to bottom in the middle range of $\delta$ corresponds to L-W, 
L-HS, NL-SH, TM1, NLC, and NL1; the dashed curve corresponds to 
Myers-Swiatecki' result. One source of deviation among these curves 
comes from the difference in the origin of the curves: 
$\rho_0=\rho_m(0)$. In cases of L-W and L1 $\rho_0$ is much higher but 
others are close or lower than that of Myers-Swiatecki's. However, even 
if all the curves are rescaled to the same $\rho_0$, there still exists 
large deviation among these curves in the middle range of $\delta$. 

Fig.4 plots some examples of $e_m${\it versus} $\delta$, where the solid 
curve from left to right on the end of the curve corresponds to L-HS, 
NL-SH, TM1, NL1, NLC, and L-W; the dashed curve corresponds to 
Myers-Swiatecki' result. All curves are close each other in the low 
asymmetry region, but L-W's is significantly lower than others for 
$\delta>0.2$.

Fig.5 is the curve $K_m$ {\it versus} $\delta$ calculated by same 
parameter sets as that of Figs.3 and 4. The solid curve from top to 
bottom in the middle range of $\delta$ is by L-W, L-HS, NL-SH, TM1, 
NLC, and NL1, and the dashed curve is by Myers-Swiatecki. The diference 
between these curves is obvious, even if NLC and NL1's are close to each other 
as well as  close to Myers-Swiatecki's.

\vskip 0.5cm
\centerline{\bf 6. Summary }
\vskip 0.5cm

In summary, the properties of nuclear matter at standard density 
$\rho_0$ with equal neutron and proton densities,  $\rho_n=\rho_p$, are 
calculated at first in the relativistic mean field theory with a variety 
of parameter sets. The result shows that the volume energy $a_1$ and 
symmetry energy $J$ are around the acceptable value $16$MeV and $30$MeV 
respectively, the incompressibility $K_0$ is reasonable only for 
nonlinear model while is unacceptably high for linear model, the density 
symmetry $L$ is around $100$MeV for most parameter sets, and the 
symmetry incompressibility $K_s$ has positive value whose sign is 
opposite to most expectations based on the nonrelativistic model.

Secondly, the calculation shows that for most parameter sets  there 
exists a critical point $(\rho_c, \delta_c)$, where the minimum and the 
maximum of the equation of state are coincident and the 
incompressibility equals zero, and it falls into ranges 
$0.014$fm$^{-3}<\rho_c<0.039$fm$^{-3}$ and $0.74<\delta_c\le0.95$; while 
for some parameter sets there is no critical point and the pure neutron 
matter is bound. The deviation among results calculated by different 
parameter sets is discussed. The maximum mass of neutron stars is also 
calculated with results in the range 
$2.45$M$_\odot\leq M_{NS}\leq 3.26$M$_\odot$. It is worthwhile to note 
that a more realistic calculation, by using a nuclear Thomas-Fermi 
equation of state, gives a maximum mass of neutron stars equal to 
$3.26$M$_\odot$\cite{Chung78}. The most of observational neutron star 
masses are between $1.2-1.8$M$_\odot$.

As different parameter sets give results which deviate significantly 
from one another, in order to extract from them more reliable predictions 
for nuclear matter properties, more sophisticated data fit, especially 
the data fit to larger number of nuclear masses and other measured 
nuclear data is expected for the nonlinear $\sigma$-$\omega$-$\rho$ 
model of relativistic mean field theory.

\appendix
\section{}

Functions $F_m(x)$ and $f_m(x)$ defined below are useful in the 
analytical expressions and numerical calculations of the relativistic 
mean field theory:
\begin{equation}F_m(x)\equiv\int_0^xdx\cdot 
x^{2m}\sqrt{1+x^2},\,\,\,\,m\ge 1,
\end{equation}
\begin{equation}f_m(x)\equiv \int_0^xdx\frac{x^{2m}}{\sqrt{1+x^2}},
\,\,\,\,m\ge 1.
\end{equation}
The following formulas can be obtained:
\begin{equation}F_m(x)=f_m(x)+f_{m+1}(x),
\end{equation}
\begin{equation}f_{m+1}'(x)=x^2f_m'(x),
\end{equation}
\begin{equation}F_{m+1}'(x)=x^2F_m'(x),
\end{equation}
\begin{equation}F_m'(x)=(1+x^2)f_m'(x),
\end{equation}
\begin{equation}f_m(x)=xF_{m-1}'(x)-(2m-1)F_{m-1}(x),
\end{equation}
\begin{equation}f_m(x)=-xF_m'(x)+2(m+1)F_m(x).
\end{equation}
Some examples of $F_m(x)$ and $f_m(x)$ are:
\begin{equation}
F_1(x)=\frac 18[(1+2x^2)x\sqrt{1+x^2}+ln(\sqrt{1+x^2}-x)],
\end{equation}
\begin{equation}f_1(x)=\frac 12[x\sqrt{1+x^2}+ln(\sqrt{1+x^2}-x)],
\end{equation}
\begin{equation}f_2(x)=-\frac 38\Big[\Big(1
-\frac 23x^2\Big)x\sqrt{1+x^2}+ln(\sqrt{1+x^2}-x)\Big].
\end{equation}
For $x\ll 1$, we have
\begin{equation}F_m(x)=\frac{x^{2m+1}}{2m+1}+\frac{x^{2m+3}}{2(2m+3)}
-\frac{x^{2m+5}}{8(2m+5)}+\cdots,
\end{equation}
\begin{equation}f_m(x)=\frac{x^{2m+1}}{2m+1}-\frac{x^{2m+3}}{2(2m+3)}
+\frac{3x^{2m+5}}{8(2m+5)}+\cdots.
\end{equation}

\begin{figure}
\caption{Composite  parameters $C_\sigma^2$ and $C_\omega^2$ as a function 
of $r_0$ for given $a_1=16$MeV, in the linear $\sigma$-$\omega$-$\rho$ 
model.}
\label{Figure1}
\end{figure}

\begin{figure}
\caption{The nuclear matter incompressibility $K_0$ calculated as 
a function of $a_1$ for given $r_0=1.14$fm, in the linear 
$\sigma$-$\omega$-$\rho$ model.}
\label{Figure2}
\end{figure}

\begin{figure}
\caption{Some examples of the location $\rho_m(\delta)$ of the 
$\sigma$-$\omega$-$\rho$ model equation of state. The solid curves from 
top to bottom in the middle range of $\delta$ correspond to L-W, L-HS, 
NL-SH, TM1, NLC, and NL1, respectively. The dashed curve corresponds to 
Myers-Swiatecki's result.}
\label{Fifure3}
\end{figure}

\begin{figure}
\caption{Some examples of the deep $e_m\times\delta$ of the 
$\sigma$-$\omega$-$\rho$ model equation of state. The solid curves from 
left to right on the high $\delta$ range correspond to L-HS, NL-SH, 
TM1, NL1, NLC, and L-W, respectively. The dashed curve corresponds to 
Myers-Swiatecki's result.}
\label{Fifure4}
\end{figure}

\begin{figure}
\caption{The curve $K_m\times\delta$ calculated by the same 
$\sigma$-$\omega$-$\rho$ model parameter sets as shown in Figs.4 and 5. 
The solid curves from top to bottom in the middle range of $\delta$ 
correspond to L-W, L-HS, NL-SH, TM1, NLC, and NL1, respectively, and the 
dashed curve corresponds to Myers-Swiatecki's result.}
\label{Fifure5}
\end{figure}

\begin{table}[h]
\caption{Some parameter sets of the $\sigma$-$\omega$-$\rho$ model in the 
relativistic mean-field theory. See text for details.}
\begin{tabular}{lrrrrrrrrrr}
Set &$M$&$m_\sigma$&$m_\omega$&$m_\rho$&$g_\sigma$&$g_\omega$ &$g_\rho$&$g_2$&$g_3$&$c_3$ \\
\hline
L-W  &939.0 &550.000 &783.000 &763. & 9.57269 &11.67114 &.00000  &.00000 &.0000 &.0000 \\
L-HS &939.0 &520.000 &783.000 &770. &10.47026 &13.79966 &4.03814 &.00000 &.0000 &.0000 \\
L1  &938.0 &550.000 &783.000 &763. &10.29990 &12.59990 &.00000  &.00000 &.0000 &.0000 \\
L2  &938.0 &546.940 &780.000 &763. &11.39720 &14.24780 &.00000  &.00000 &.0000 &.0000 \\
L3  &938.0 &492.260 &780.000 &763. &10.69200 &14.87050 &.00000  &.00000 &.0000 &.0000 \\
L-Z  &938.9 &551.310 &780.000 &763. &11.19330 &13.82560 &5.44415 &.00000 &.0000 &.0000 \\
NL1 &938.0 &492.250 &795.359 &763. &10.13770 &13.28460 &4.97570 &12.17240 &-36.2646 &.0000 \\
NL2 &938.0 &504.890 &780.000 &763. &9.11122 &11.49280 &5.38660 &2.30404 &13.7844 &.0000 \\
NL3 &939.0 &508.194 &782.501 &763. &10.21700 &12.86800 &4.47400 &10.43086 &-28.8849 &.0000 \\
NL3-II&939.0 &507.680 &781.869 &763. &10.20200 &12.85400 &4.48000 &10.39100 &-28.9390 &.0000 \\
NLB &939.0 &510.000 &783.000 &770. &9.69588 &12.58890 &4.27200  &2.02714 &1.6667 &.0000 \\
NL-B1 &938.9 &470.000 &783.000 &770. &8.75834 &11.80520 &3.75195 &7.51446 &-16.8112 &.0000 \\
NL-B2 &938.9 &485.000 &783.000 &770. &9.72687 &12.89370 &3.52938 &9.47080 &-28.1254 &.0000 \\
NLC &939.0 &500.800 &783.000 &770. &9.75244 &12.20370 &4.32984 &12.66960 &-33.3333 &.0000 \\
NLD &939.0 &476.700 &783.000 &770. &8.26559 &10.86600 &4.49305  &3.79970 &8.3333 &.0000 \\
NL-RA &939.0 &515.000 &782.600 &763. &9.62661 &11.90390 &4.52418 &8.06582 &-16.3173 &.0000 \\
NL-SH &939.0 &526.059 &783.000 &763. &10.44400 &12.94500 &4.38300 &6.90990 &-15.8337 &.0000 \\
NL-VT &938.9 &483.420 &780.000 &763. &9.79084 &12.65660 &4.61319 &13.16500 &-38.1282 &.0000 \\
NL-Z &938.9 &488.670 &780.000 &763. &10.05530 &12.90860 &4.84944 &13.50720 &-40.2243 &.0000 \\
TM1 &938.0 &511.198 &783.000 &770. &10.02890 &12.61390 &4.63220 &7.23250 &.6183 &71.3075 \\
TM2 &938.0 &526.443 &783.000 &770. &11.46940 &14.63770 &4.67830 &4.44400 &4.6076 &84.5318 \\
\end{tabular}
\label{TableI}
\end{table}

\begin{table}[h]
\caption{Standard nuclear matter properties given by the 
$\sigma$-$\omega$-$\rho$ model parameter sets listed in Table I. 
$a_1$, $K_0$, $J$, $L$ and $K_s$ are in MeV, $\rho_0$ in fm$^{-3}$. 
See text for details.}
\begin{tabular}{lrrrrrrr}
 Set &$\rho_0$&$a_1$&$K_0$ & $J$ & $L$  & $K_s$  \\
\hline
L-W   &.1937 &15.75 &545.6 &22.11 & 74.5 &  74.8 \\
L-HS  &.1485 &15.75 &546.8 &34.98 &115.5 &  93.4 \\
L1    &.1766 &18.52 &625.6 &21.68 & 75.6 &  81.8 \\
L2    &.1417 &16.78 &578.5 &19.07 & 68.8 &  97.4 \\
L3    &.1344 &18.24 &624.5 &18.86 & 69.5 & 102.1 \\
L-Z   &.1494 &17.07 &586.3 &48.84 &157.9 &  94.2 \\
NL1   &.1518 &16.42 &211.1 &43.46 &140.1 & 142.6 \\
NL2   &.1456 &17.03 &399.4 &43.86 &129.7 &  20.1 \\
NL3   &.1482 &16.24 &271.6 &37.40 &118.5 & 100.8 \\
NL3-II&.1491 &16.26 &271.7 &37.70 &119.7 & 103.3 \\
NLB   &.1485 &15.77 &421.0 &35.01 &108.3 &  54.8 \\
NL-B1 &.1625 &15.79 &280.4 &33.04 &102.5 &  76.1 \\
NL-B2 &.1627 &15.79 &245.6 &33.10 &111.3 & 158.8 \\
NLC   &.1485 &15.77 &224.4 &35.02 &108.0 &  76.8 \\
NLD   &.1485 &15.77 &343.2 &35.01 &101.5 &  13.5 \\
NL-RA &.1570 &16.25 &320.5 &38.90 &119.1 &  62.0 \\
NL-SH &.1460 &16.35 &355.3 &36.12 &113.6 &  79.7 \\
NL-VT &.1530 &16.09 &172.8 &39.73 &126.9 & 130.0 \\
NL-Z  &.1508 &16.19 &172.8 &41.72 &133.9 & 140.0 \\
TM1   &.1452 &16.26 &281.2 &36.89 &110.8 &  33.5 \\
TM2   &.1323 &16.16 &343.8 &35.98 &113.0 &  56.0 \\
$\langle$NL$\rangle$ &.1500&16.14&287.7&37.53&117.1&83.2\\
MS    &.1611 &16.24 &234.4 &32.65 & 49.9 &-147.1 \\
\end{tabular}
\label{TableII}
\end{table}

\begin{table}[h]
\caption{Nuclear matter properties at the critical point 
$(\rho_c,\delta_c)$ or $(\rho_m,1)$, the maximum neutron star mass 
$M_{NS}$ and the corresponding star radius $R_{NS}$, calculated by the 
$\sigma$-$\omega$-$\rho$ model parameter sets listed in Table I. 
$\rho_c$ is in fm$^{-3}$, $e_c$ and $K_c$ in MeV, $M_{NS}$ in solar mass 
M$_\odot$ and $R_{NS}$ in km. Myers-Swiatecki's values are listed in 
the last row for comparison. See text for details.}
\begin{tabular}{lrrrrrrr}
Set&$\delta_c$ &$\rho_c$ &$M^*/M$ &$e_c$ &$K_c$ &$M_{NS}$ &$R_{NS}$ \\
\hline
L-W   &1.00& .0987& .766&  1.93&  77.8& 2.60 &12.2 \\
L-HS  & .86& .0392& .872&  2.75&   0.0& 3.08 &14.6 \\
L1    &1.00& .1034& .718& -0.63& 142.3& 2.80 &13.0 \\
L2    &1.00& .0849& .712& -1.04& 138.4& 3.13 &14.4 \\
L3    &1.00& .0847& .688& -2.46& 174.5& 3.26 &15.0 \\
L-Z   & .75& .0388& .871&  2.51&   0.0& 3.16 &15.1 \\
NL1   & .91& .0150& .951&  1.43&   0.0& 2.96 &14.2 \\
NL2   & .81& .0288& .925&  2.41&   0.0& 2.78 &13.9 \\
NL3   & .92& .0182& .943&  1.62&   0.0& 2.91 &13.9 \\
NL3-II& .92& .0178& .944&  1.62&   0.0& 2.91 &13.9 \\
NLB   & .87& .0327& .906&  2.43&   0.0& 2.87 &13.8 \\
NL-B1 & .95& .0236& .936&  1.94&   0.0& 2.68 &12.9 \\
NL-B2 &1.00& .0212& .934&  1.75&   1.8& 2.87 &13.5 \\
NLC   & .95& .0175& .949&  1.63&   0.0& 2.77 &13.2 \\
NLD   & .87& .0302& .929&  2.29&   0.0& 2.60 &13.0 \\
NL-RA & .87& .0243& .934&  1.89&   0.0& 2.75 &13.4 \\
NL-SH & .90& .0235& .927&  1.90&   0.0& 2.93 &14.1 \\
NL-VT & .95& .0151& .952&  1.44&   0.0& 2.87 &13.7 \\
NL-Z  & .94& .0144& .953&  1.39&   0.0& 2.92 &13.9 \\
TM1   & .90& .0217& .935&  1.82&   0.0& 2.45 &13.3 \\
TM2   & .90& .0217& .918&  1.83&   0.0& 2.73 &14.4 \\
MS    & .82& .0304&     &  1.10&   0.0&      &     \\   
\end{tabular}
\label{TableIII}
\end{table}

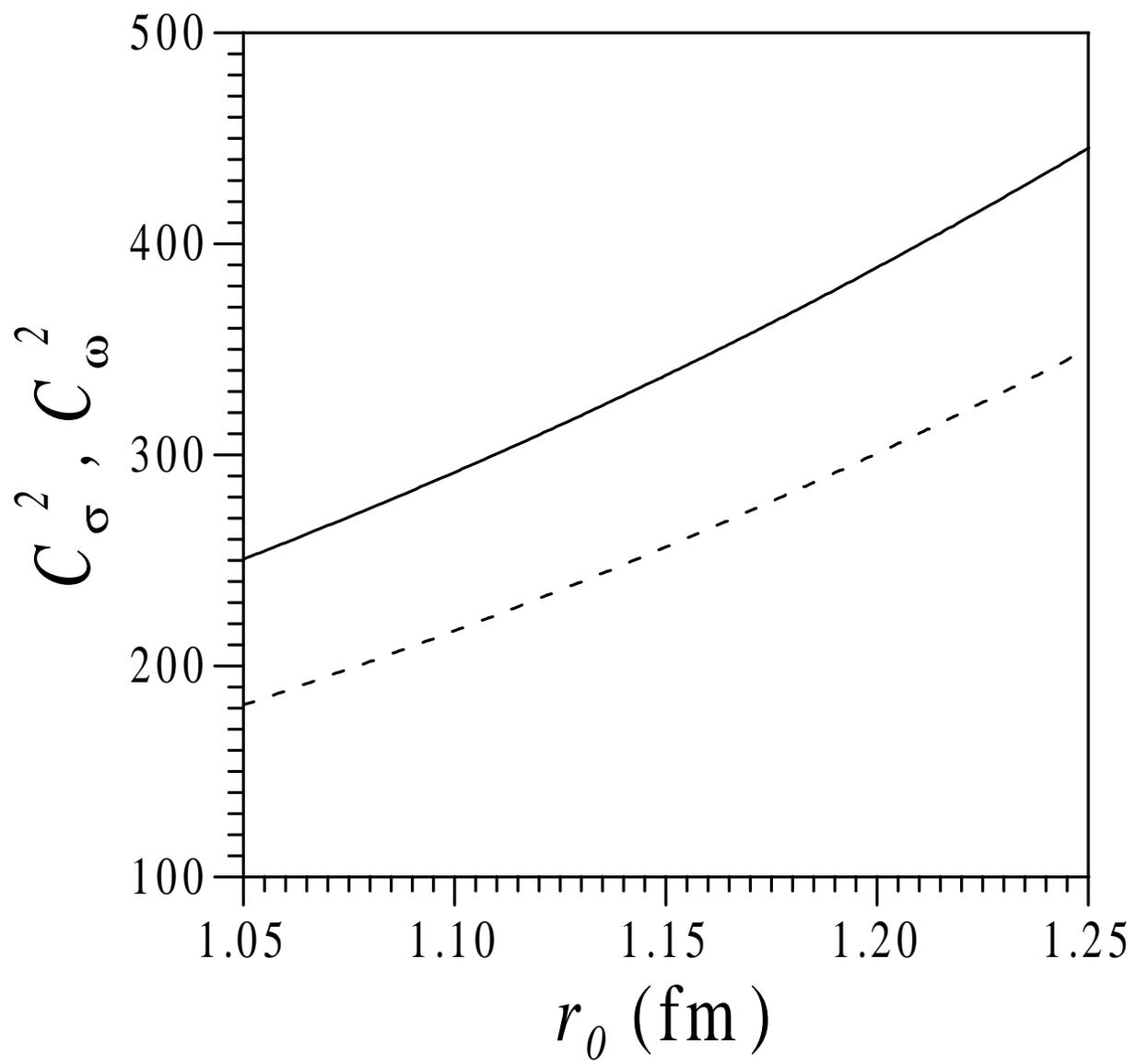
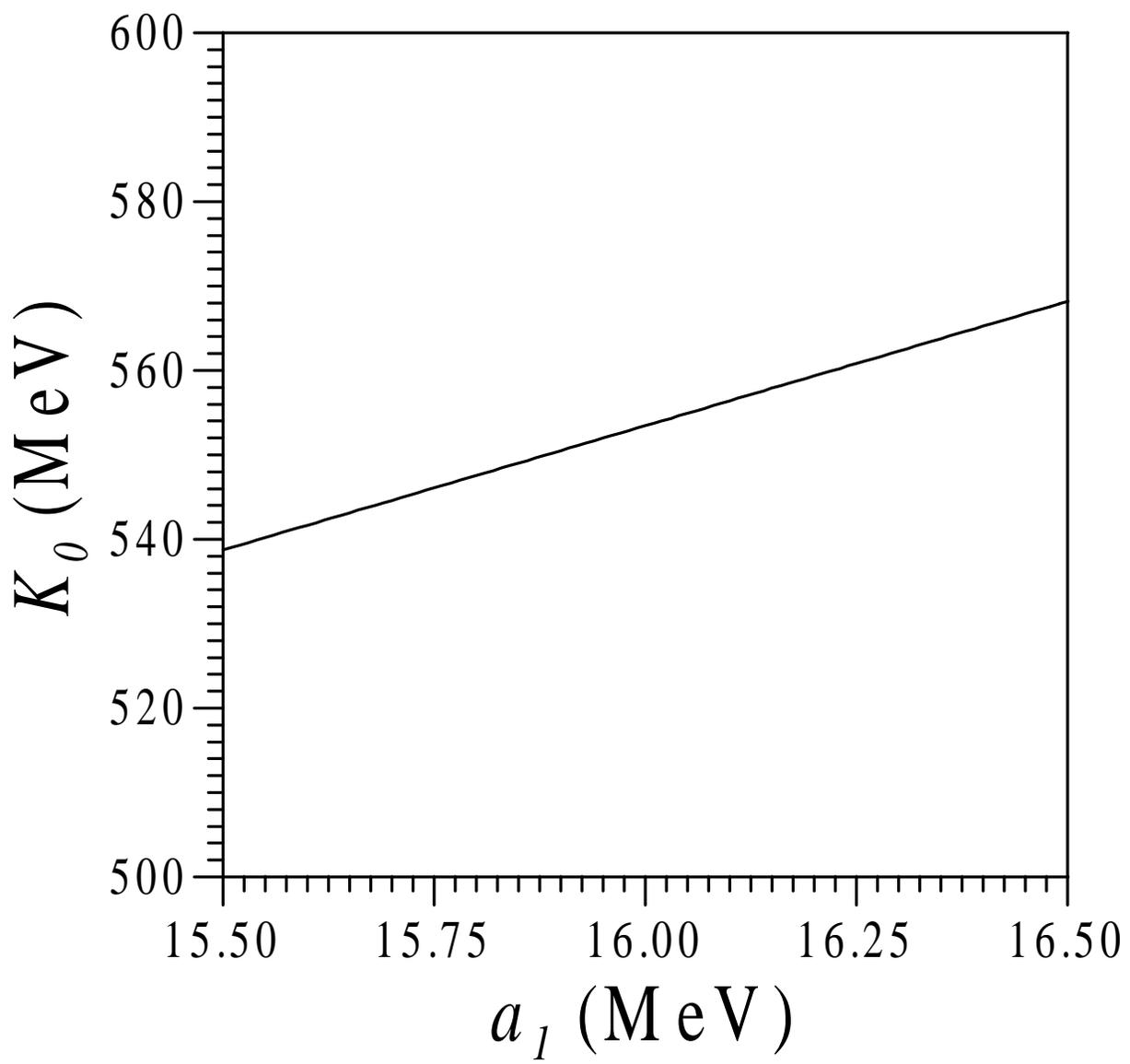
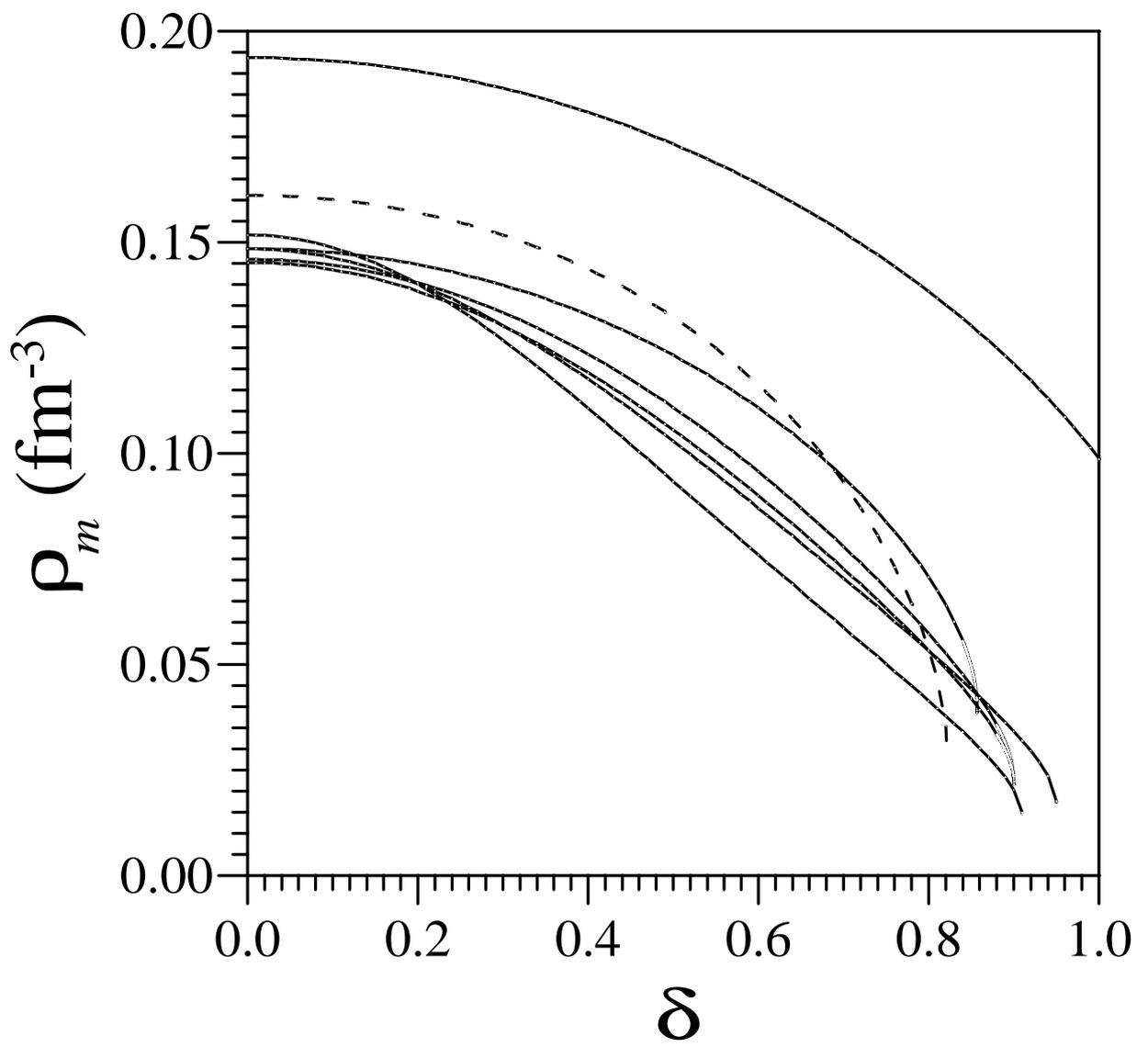
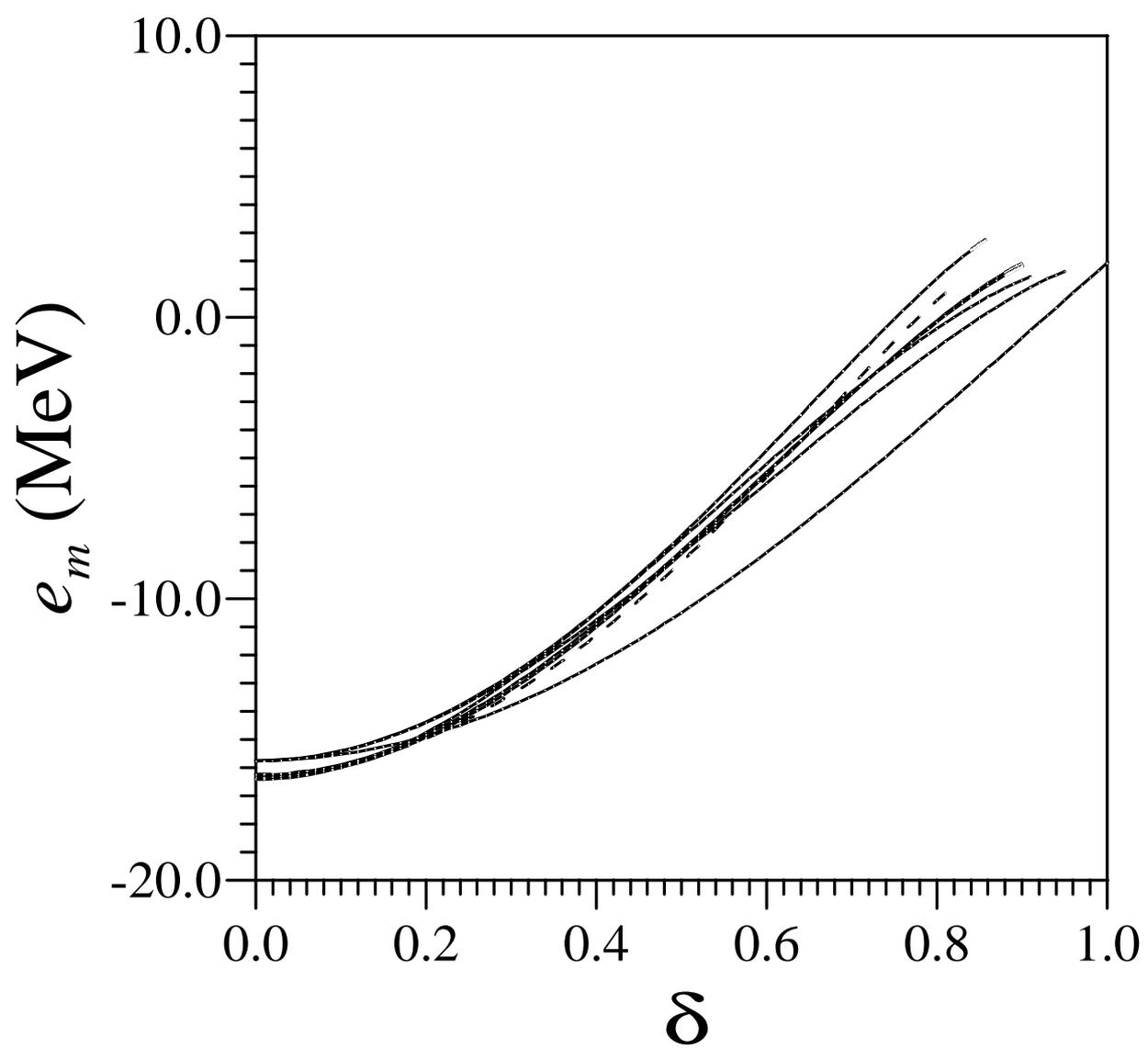
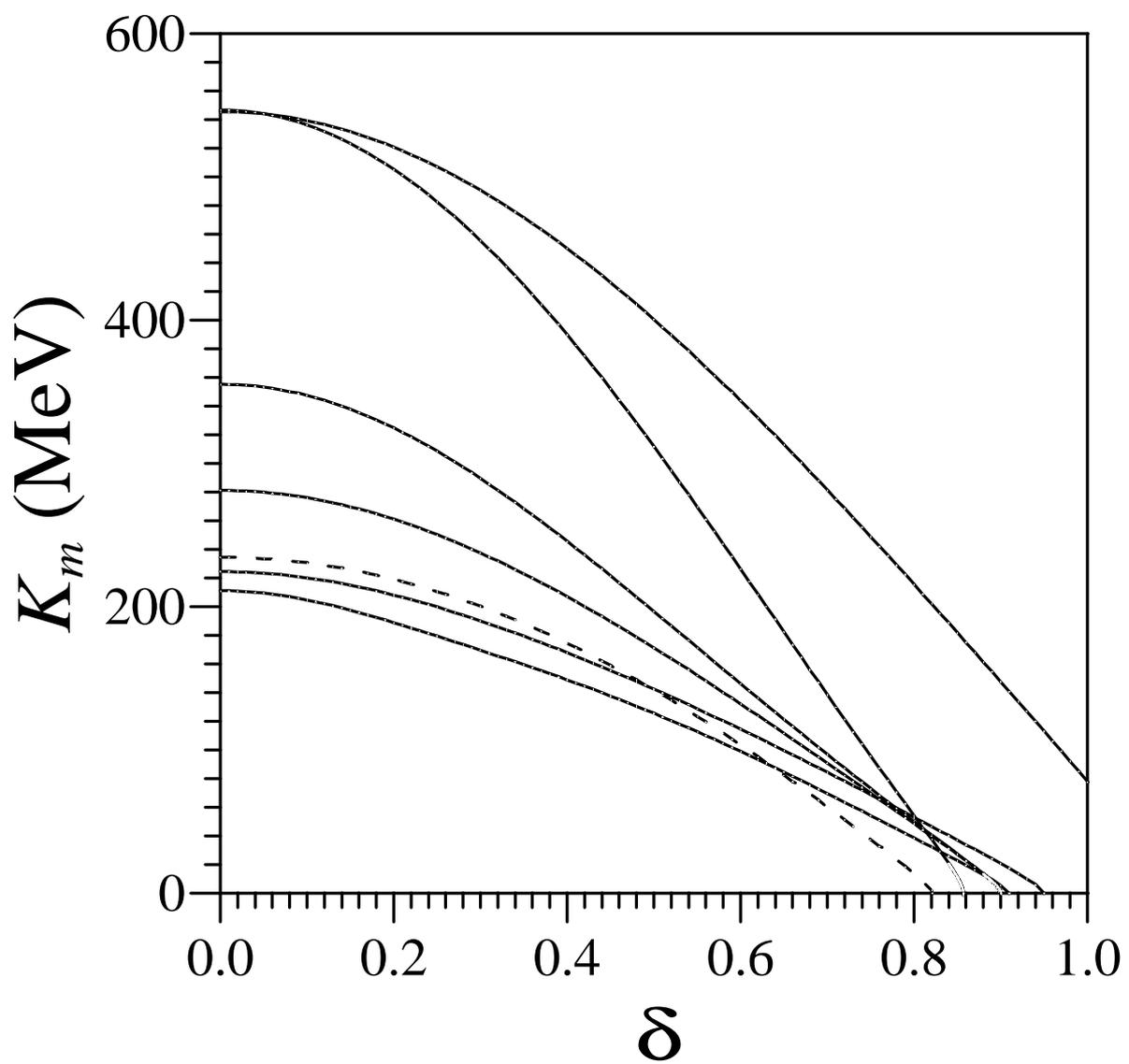
\end{document}